\documentclass{mem}
\usepackage{natbib}\usepackage{txfonts}\usepackage{balance}
\usepackage{graphicx}
\usepackage[a4paper]{hyperref}
\idline{75}{282}
\begin{document}
\def\teff{$T\rm_{eff }$}
\def\kms{$\mathrm {km s}^{-1}$}

\title{
A statistical investigation of the radio emission of clusters : role of AGNs
}

   \subtitle{}

\author{
J. Lanoux\inst{1,2}, E. Pointecouteau\inst{1,2} 
\and M. Giard\inst{1,2}
          }

  \offprints{J. Lanoux}

\institute{
(1) Universite de Toulouse (UPS-OMP), Institut de Recherche en Astrophysique et Planetologie\\
(2) CNRS, UMR 5277, 9 Av. colonel Roche, BP 44346, F 31028 Toulouse cedex 4, France\\
\email{lanoux@cesr.fr}
}

\authorrunning{Lanoux}

\titlerunning{A statistical investigation of clusters radio emission : role of AGNs}

\abstract{
Radio emission in the direction of galaxy clusters is due to the individual emission of AGNs, and to the diffuse non-thermal emission of the intracluster medium (ICM). The population of AGNs in clusters is correlated to the overall halo properties, and likely impacts their evolutions. In order to better understand this connection (which will also leads to constraints on the non-thermal ICM emission for lower mass systems), we are conducting a statistical analysis of the radio emission in the direction of a large number of X-ray clusters. By means of their stacked radio emission, we are investigating the radio luminosities in clusters of galaxies with respect to their mass, with the goal to better understand their link to and their imprint on the intra-cluster gas physics.

\keywords{Radio: galaxies: clusters -- X-rays: galaxies: clusters -- Radio: galaxies -- Intergalactic medium -- Diffuse radiation -- Cosmology: observations }
}

\maketitle{}

\section{Introduction}

The ICM is mainly constituted of diffuse hot gas that cools down via thermal Bremsstrahlung emission observed at X-ray wavelengths. But it is also populated by non-thermal relativistic electrons observed at radio wavelengths due do their synchrotron emission. The radio non-thermal emission of galaxy clusters is due to point-like sources (AGNs, radio galaxies) and diffuse extended sources (halos, relics, mini-halos, e.g. \citealp{ferrari} for a review). 

Diffuse radio emission has now been observed in a few tens of massive merging galaxy clusters. \citet{liang} were the first to show the existing correlation between radio and X-ray luminosities in such clusters. This correlation was confirmed by subsequent works such as \citet{giovannini} or \citet{brunetti}.

Here, we investigated this correlation down to a lower mass regime of clusters, together with the contribution of AGNs to the total radio luminosity. So far, we have been limited by the sensitivity of radio surveys to directly detect diffuse emission. 

Therefore, we investigated the correlations by means of statistical and stacking tools applied to X-ray (RASS) and radio (NVSS) all-sky surveys. We used the Meta-Catalogue of X-ray detected Clusters of galaxies (MCXC, 1743 clusters, \citealp{piffaretti}). From maps extracted at each cluster position, we computed count rates and fluxes within $R_{500}$. We derived corresponding X-ray and radio luminosities.  From the NVSS source catalogue \citep{condon} and the 2MASS extended source catalogue \citep{jarrett}, we selected AGNs associated with clusters and we derived their summed radio luminosities within $R_{500}$ for each cluster. We then fitted the $L_{R} - L_X$ correlations with BCES regression method \citep{akritas}.

\section{Results}

\begin{figure}[]
\includegraphics[scale=0.36]{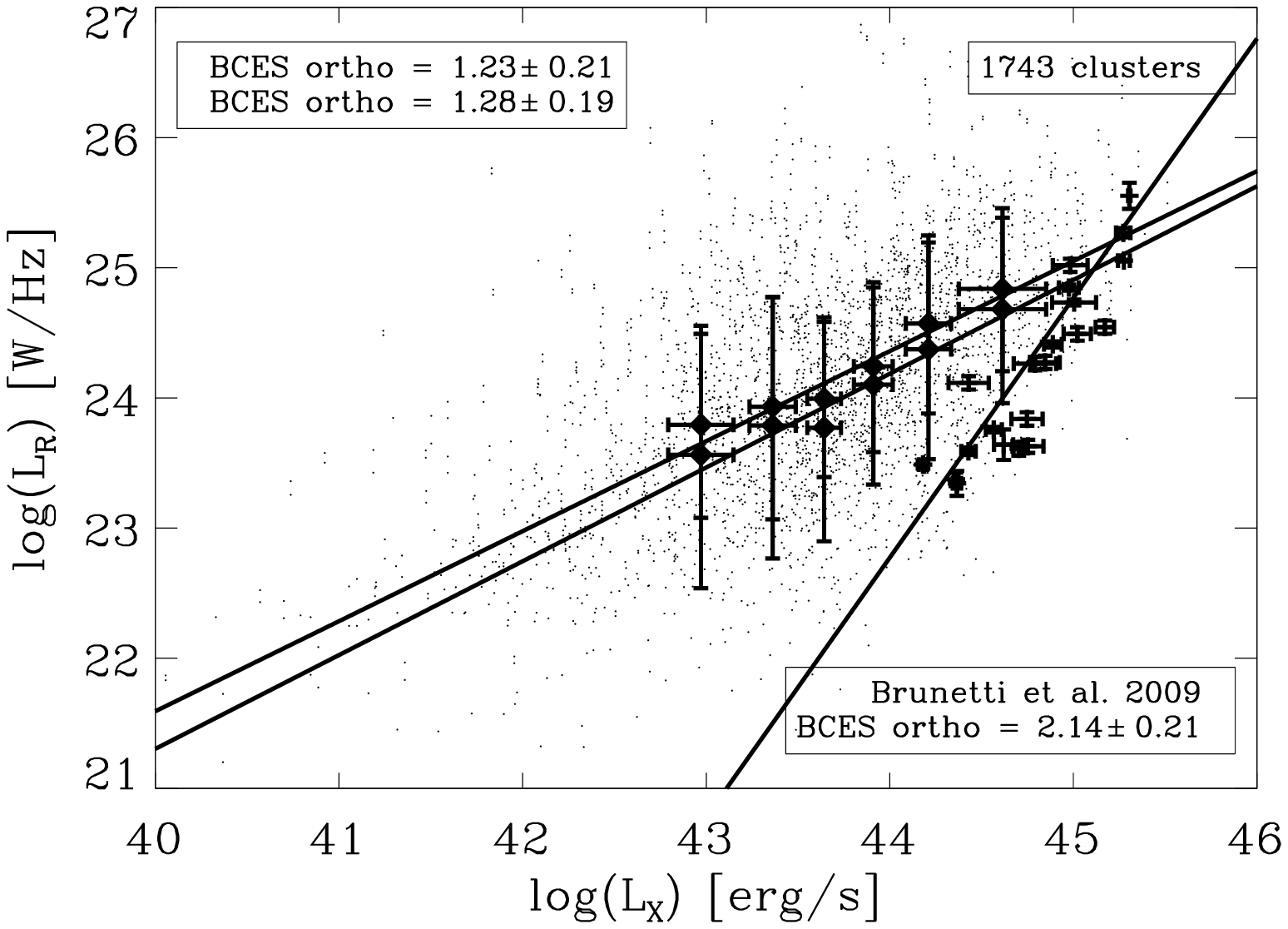}
\includegraphics[scale=0.36]{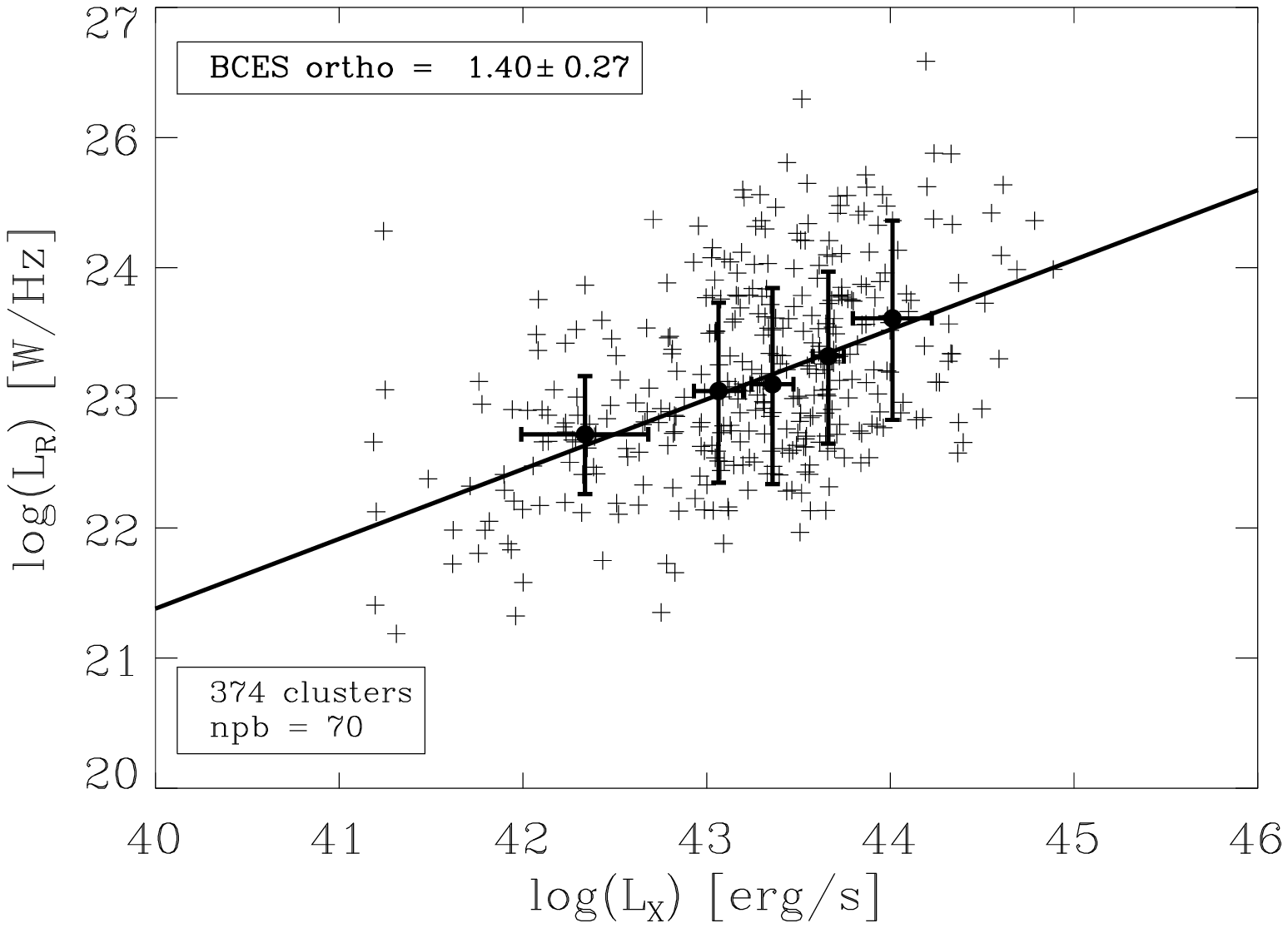}
\caption{
\footnotesize
$L_{R} - L_X$ correlations for total integrated radio luminosities within $R_{500}$ and for luminosities of point sources only (top). $L_{R} - L_X$ correlation for radio luminosities of AGNs for a sub-sample of 374 clusters with redshift $z < 0.1$ (bottom).}
\label{fig}
\end{figure}

We found a $L_{R} - L_X$ correlation with a slope of $\sim$1.2 for total integrated radio luminosities within $R_{500}$ (Figure \ref{fig}, top). This slope is flatter compared to other works which investigated the relation between detected diffuse radio emissions and $L_X$ (i.e. $\sim$2.1 for \citealp{brunetti} or $\sim$1.7 for \citealp{giovannini}). 

We found that the summed radio luminosities of selected point sources within $R_{500}$ compares very well to the total integrated radio luminosities within this aperture. This likely shows that the component of point sources (i.e. AGNs) is dominant in our aperture. 

This conclusion is reinforced by the strong $L_R - L_X$ correlation we found for radio luminosities associated to AGNs only for a sub-sample of 374 galaxy clusters of the MCXC with redshift $z < 0.1$ (Figure \ref{fig}, bottom). We derive a slope of $\sim$1.4 compatible with the previous correlation found for total integrated radio luminosities. This strengthened our result for a flatter $L_{R} - L_X$ with respect to the correlation found for diffuse radio emission in clusters.

\section{Conclusion and Perspectives}

We have probe the radio emission of a large number of X-ray clusters, and we have shown that there is a strong  $L_{R} - L_X$ correlation, and that the radio luminosity is mainly contributed by AGNs in clusters.

The provided statistical constraints will allow us to further investigate the overall nature of the radio emission seen in the direction of galaxy clusters.

%\begin{acknowledgements}
%\end{acknowledgements}

\bibliographystyle{aa}

\end{document}